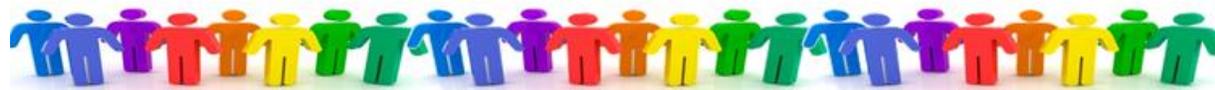

# Does Preregistration Improve the Credibility of Research Findings?

Mark Rubin

*The University of Newcastle, Australia*

## Abstract

Preregistration entails researchers registering their planned research hypotheses, methods, and analyses in a time-stamped document before they undertake their data collection and analyses. This document is then made available with the published research report to allow readers to identify discrepancies between what the researchers originally planned to do and what they actually ended up doing. This *historical transparency* is supposed to facilitate judgments about the credibility of the research findings. The present article provides a critical review of 17 of the reasons behind this argument. The article covers issues such as HARKing, multiple testing, *p*-hacking, forking paths, optional stopping, researchers' biases, selective reporting, test severity, publication bias, and replication rates. It is concluded that preregistration's historical transparency does not facilitate judgments about the credibility of research findings when researchers provide *contemporary transparency* in the form of (a) clear rationales for current hypotheses and analytical approaches, (b) public access to research data, materials, and code, and (c) demonstrations of the robustness of research conclusions to alternative interpretations and analytical approaches.

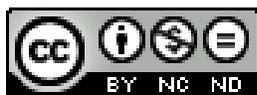

Correspondence concerning this article should be addressed to Mark Rubin at the School of Psychology, Behavioural Sciences Building, The University of Newcastle, Callaghan, NSW 2308, Australia.
E-mail: Mark.Rubin@newcastle.edu.au  Web: http://bit.ly/rubinpsyc

Preregistration has recently become popular in psychology and other disciplines. Preregistration entails researchers registering their planned research hypotheses, methods, and analyses in a time-stamped document before they undertake their data collection and analyses (Nosek et al., 2018; Wagenmakers et al., 2012). This document is then made available with the published research report to allow readers to identify discrepancies between what the researchers originally planned to do and what they actually ended up doing. Knowledge of these discrepancies is supposed to facilitate readers' judgments about the credibility of research findings (e.g., Centre for Open Science, n.d.; Nosek et al., 2019; Nosek et al., 2018; Wagenmakers et al., 2012). As Nosek et al. (2018) concluded in their *Preregistration Revolution* article, "preregistration improves the interpretability and credibility of research findings" (p. 2605). But is this really the case?[1] To address this question, I distinguish between two forms of transparency: *historical transparency* and *contemporary transparency*.

Historical transparency allows readers to identify differences between a researcher's current hypotheses, methods, and analyses and their planned hypotheses, methods, and analyses. In contrast, contemporary transparency allows readers to identify differences between a researcher's current hypotheses, methods, and analyses and a range of other potential hypotheses, methods, and analyses, regardless of whether or not those other hypotheses, methods, and analyses were originally planned by the researcher. Historical transparency is necessary in order for readers to judge whether a researcher's current approach is better or worse than their planned approach. Contemporary transparency is necessary in order for readers to judge whether a researcher's current approach is better or worse than a range of other potential approaches that may or may not include their planned approach.

Preregistration only provides historical transparency. It does not provide contemporary transparency. Contemporary transparency may be provided by (a) clear rationales for current hypotheses and analytical approaches, (b) public access to research data, materials, and code, and (c) demonstrations of the robustness of research conclusions to alternative interpretations and analytical approaches.

In the present article, I question whether preregistration's historical transparency facilitates judgments of the credibility of research findings over and above the information that is provided by contemporary transparency. In other words, I ask whether preregistration's historical transparency produces benefits beyond those provided by contemporary transparency.

Several previous articles have questioned the added value of preregistration (e.g., Baron, 2018; Coffman & Niederle, 2015; Devezer et al., 2020; Donkin & Szollosi, 2020; Ledgerwood, 2018; Lewandowsky, 2019; Navarro, 2019; Oberauer & Lewandowsky, 2019; Szollosi & Donkin, 2019; Szollosi et al., 2020). However, none of these articles have distinguished between historical and contemporary transparency, and none have considered the full range of advantages that are supposed to be associated with preregistration.

In the present article, I provide a critical review of 17 of the proposed benefits of preregistration. The benefits are grouped according to the three key stages of hypothesis testing: (1) generating hypotheses, (2) data analysis, and (3) research conclusions. Based on this review, I conclude that preregistration's historical transparency does not facilitate judgments about the credibility of research findings over and above the information provided by contemporary transparency.

# Generating Hypotheses

Preregistration can help researchers to develop high quality hypotheses and hypothesis tests. Preregistration's historical transparency can also inform readers about (a) whether researchers generated their hypotheses before or after conducting their tests and (b) any a priori hypotheses that have not been addressed in the research report. I discuss each of these issues in turn.

## Facilitating the Development of Hypotheses and Their Tests

Preregistration can help researchers to carefully plan their hypotheses and hypothesis tests (Field et al., 2020). In addition, registered reports allow researchers to receive peer review feedback about their research approach prior to data collection (Chambers et al., 2019; Field et al., 2020). This planning and feedback may improve the quality of research projects. However, neither planning nor feedback are essential features of preregistration (Lakens, 2019, p. 223). Hence, preregistered research projects can be poorly planned and have received no peer review feedback. In contrast, non-preregistered research projects can be carefully planned and extensively peer reviewed (e.g., as might occur in the case of a research grant application). Hence, planning and feedback do not distinguish between preregistered and non-preregistered research. Consequently, although planning and feedback may improve the quality of research projects, they are not essential features of preregistration, and so preregistration does not necessarily confer an advantage in this respect.

## Identifying HARKing

HARKing refers to the practice of undisclosed *hypothesizing after the results are known* (Kerr, 1998). In this case, researchers conduct a test, observe their test result, conceive or retrieve a hypothesis that is relevant to this test result, and then present this hypothesis in their research report as if it was an a priori hypothesis rather than a post hoc hypothesis.

Preregistration's historical transparency can be used to detect HARKing. If hypotheses are included in a preregistration document, then *hypothesis generation* must have occurred before *hypothesis testing* (e.g., Centre for Open Science, n.d.; Nosek et al., 2018; Wagenmakers et al., 2012).

However, historical transparency is not necessary to detect HARKing. Contemporary transparency may also reveal HARKing. Specifically, researchers may indicate in their research report that their hypotheses were generated after they became aware of their current results. Hollenbeck and Wright (2017) have described this research practice as *transparently hypothesising after the results are known* or THARKing.

It is also important to appreciate that information about the time at which hypotheses are deduced from objectively a priori theory and evidence does not affect the credibility of hypothesis tests (Lewandowsky, 2019; Mayo. 2015, p. 61; Oberauer & Lewandowsky, 2019; Rubin, 2017c, 2019a; Vancouver, 2018; Worrall, 2014). I provide two common examples.

First, a researcher who has viewed their current results may retrieve an old hypothesis from previously published work and then claim that hypothesis to be supported or contradicted by their current results (RHARKing, Rubin, 2017c, pp. 314-316). For example, Bleidorn et al. (2016, p. 396) proposed that "men tend to have higher self-esteem than women." Hence, any researcher who has measured gender and self-esteem in their study can undertake an unplanned test of this objectively a priori hypothesis. In this case, the researcher's current result may be used to increase or decrease support for the hypothesis because the information provided by the result is *epistemically independent* from the information that forms the basis for the hypothesis (Rubin, 2019a). Hence, the result is *use novel* with respect to the hypothesis, and there is no *double counting* (Mayo, 1996; Worrall, 2014). All that is left to do is to appraise the validity with which the RHARKed hypothesis has been tested, and this appraisal can be made by readers without knowing that the hypothesis test was unplanned and that the researcher retrieved the hypothesis after knowing their test result.

Second, a researcher may be inspired by their current result to deduce a new hypothesis from objectively a priori theory and evidence that accounts for that result. In this case, they may provide a theoretical rationale in their research report that explains the steps in this deduction from theory to hypothesis. Again, the current result can be shown to be epistemically unnecessary in this theoretical rationale and, therefore, use novel with respect to the hypothesis. Consequently, readers can (a) form an initial belief about the plausibility of the hypothesis that is based solely on the theoretical rationale and then (b) update that belief based on the additional information that is provided by the current result. Readers may undertake this process in a valid manner even if they are unaware that the researcher's current result secretly inspired their deduction of the hypothesis, because this inspiration does not form part of the information that is required in the theoretical rationale (Rubin, 2019a). For example, imagine the reader of a research report who is presented with the secretly HARKed hypothesis that "eating apples improves mood" (Rubin, 2019a). The reader may form an initial belief about the plausibility of this hypothesis based on the researcher's secretly HARKed theoretical rationale, which is deduced from objectively a priori theory and evidence that (a) Vitamin C improves mood, and (b) apples are rich in Vitamin C. The reader may then update their initial belief based on the researcher's current result that participants' mean mood increased significantly after they ate an apple a day for a week, relative to a control group.

Of course, flexible theorizing allows researchers to deduce a potentially infinite number of hypotheses from a priori theory and evidence (Kerr, 1998, p. 210), and these hypotheses can then be used to explain any set of results. However, this flexible theorizing will often result in poor quality theoretical rationales that obviously lack coherence, breadth, depth, and parsimony. These negative theoretical qualities will detract from the plausibility of the hypotheses and the credibility of the research conclusions. Importantly, poor quality theorizing can be identified and taken into account by readers in the absence of preregistration (see also Szollosi & Donkin, 2019). Hence, preregistration does not provide a credibility advantage in this respect.

## Identifying A Priori Hypotheses That Would be Otherwise Suppressed

Preregistration's historical transparency may also be used to identify researchers' a priori hypotheses that would be otherwise suppressed in their research report, possibly because they yielded null or disconfirming results (Rubin, 2017c). However, the identification of such hypotheses is not always necessary. In particular, if suppressed a priori hypotheses are irrelevant to the final research conclusions, then their suppression will not bias those conclusions (Leung, 2011; Rubin, 2017c, 2019a; Vancouver, 2018). The identification of a priori hypotheses is only necessary if those hypotheses are relevant to the reported research conclusions. Preregistration's historical transparency is not necessary to identify relevant a priori hypotheses, because their relevance makes them obvious as alternative explanations during pre- or post-publication peer review (Kerr, 1998, p. 208; Rubin, 2019a). Hence, the suppression of relevant a priori hypotheses is unlikely to be a serious problem.

Finally, contemporary transparency may be used to reveal both relevant and irrelevant hypotheses. In particular, if research materials, data, and coding information are made publicly available, then both relevant and irrelevant hypotheses can be tested by other interested researchers (Rubin, 2019a).

# Data Analysis

Preregistration is supposed to facilitate the interpretability of data analyses (e.g., Nosek et al., 2018). Here, I consider two categories of data analyses: (1) general data analyses (deviations from planned analyses and distinguishing confirmatory and exploratory analyses) and (2) data analyses involving significance testing (alpha levels, multiple testing, forking paths, $p$-hacking, optional stopping, and $p$ values in exploratory analyses).

## General Data Analyses
### Identifying Deviations from Planned Analyses

Preregistration's historical transparency may prompt researchers to provide rationales for deviating from their planned analyses (Nosek et al., 2018, p. 2602). However, as scientists rather than historians, we should be more interested in researchers' rationales for their *current* methods and analyses rather than in their rationales for *historical changes* that have led to those current method and analyses. For example, if a researcher preregisters Test A, but then decides to use Test B instead, we should be more interested in the rationale for using Test B than in the rationale for the change from Test A to Test B. Of course, the rationale for Test B may include a discussion of its superiority to a range of other similar tests, which may include Test A. However, this rationale may be put forward without knowing that the researcher originally planned to use Test A.

Contemporary transparency provides information about a researcher's current data analyses and how their results would be different in the case of alternative relevant analyses. The validity and quality of a researcher's current analyses can be checked by comparing their reported justifications and rationales for these analyses with external information (e.g., articles on research methodology and statistics) and internal information (e.g., the researcher's publicly accessible research materials, data, and coding information). The generality of a researcher's results under alternative relevant approaches can be checked via *robustness analyses* (e.g., Thabane et al., 2013; also called *sensitivity* or *multiverse* analyses) that indicate how research results vary as a function of reasonable data exclusions and alternative sensible data coding, aggregation, and analysis approaches. For example, a researcher might reveal how their pattern of significant and non-significant results varies when they conduct their analyses with and without outliers and theoretically relevant covariates. Of course, preregistered research protocols may plan to reveal this contemporary transparency. However, it is not necessary for researchers to demonstrate that they planned to demonstrate contemporary transparency in order for contemporary transparency to improve the interpretability of data analyses.

### Distinguishing Confirmatory and Exploratory Analyses

Preregistration is supposed to help to distinguish between confirmatory and exploratory data analyses, with the assumption being that the results of exploratory hypothesis tests are less credible than those of confirmatory hypothesis tests (e.g., Centre for Open Science, n.d.; Field et al., 2020; Nosek & Lindsay, 2018; Nosek et al., 2018; Wagenmakers et al., 2012). However, the distinction between confirmatory and exploratory analyses is problematic, and preregistration does not clarify it (see also Devezer et al., 2020).

Confirmatory and exploratory data analyses may be defined as analyses that are preregistered and not preregistered respectively. However, in this case, the argument that preregistration provides a credibility advantage by distinguishing between confirmatory and exploratory analyses boils down to the tautology that preregistered research is more credible than non-preregistered research because it is preregistered!

Confirmatory and exploratory analyses may also be defined in relation to a priori and post hoc hypotheses respectively. However, this distinction is problematic because the a priori/post hoc status of a hypothesis can be defined subjectively and objectively. To illustrate, imagine that a researcher conducts a test, observes their test result, and then retrieves a hypothesis from previously published literature to explain that result (i.e., RHARKing, Rubin, 2017c, pp. 314-316). Is this hypothesis post hoc or a priori? According to the subjective interpretation, the researcher only knew about the hypothesis after they conducted their test. Hence, the hypothesis is post hoc, and the test is exploratory. However, according to the objective interpretation, there is clear evidence that the hypothesis was developed and known by other people before the researcher's test. Hence, the hypothesis is a priori, and the test is confirmatory.

Preregistration provides historical transparency about the subjective knowledge of specific researchers before they undertook their analyses. Hence, it uses the subjective interpretation to distinguish between a priori and post hoc hypotheses. However, we should be more concerned about the epistemic independence between hypotheses and test results than about the order in which hypotheses and test results become subjectively known by specific researchers (Rubin, 2019a; Worrall, 2014). From this epistemic perspective, the distinction between confirmatory and exploratory analyses is irrelevant. Instead, what matters is whether hypotheses have been either (a) deduced from objectively a priori theory and evidence (prediction) or (b) induced from the current test results (accommodation). Importantly, from this epistemic perspective, prediction can occur either before or after specific researchers become subjectively aware of their test results (Rubin, 2019a; Worrall, 2014).

## Data Analyses Involving Significance Testing
### Pre-Specifying Alpha Levels

In most cases, researchers use a default, conventional alpha level (significance threshold) of $\alpha = .050$. In these cases, there is little need for researchers to justify their alpha. However, in some cases, researchers use more liberal or conservative alpha levels (e.g., $\alpha = .10$ or $\alpha = .005$). In these cases, researchers may preregister their unconventional alpha levels in order to provide evidence that their choice of alpha level was not influenced by their specific results. For example, a researcher may preregister an alpha level of .10 to provide evidence that they chose this more liberal alpha level before knowing their current results. This evidence may be particularly important if the researcher's significant results are based on $p$ values that fall in the region of .10 to .05.

However, historical transparency about the specification of unconventional alpha levels is redundant in the face of a convincing justification for those alpha levels that is supported by contemporary evidence. For example, in the case of a liberal alpha level of .10, a researcher may explain why they regard it as being less important than usual to be concerned about Type I errors given the specific circumstances of their particular significance tests. Peer reviewers and readers may then make judgments about whether or not they accept this justification. Importantly, a strong justification in the case of non-preregistered research will be more convincing than a weak justification in the case of preregistered research. Hence, preregistration's historical transparency is not necessary when contemporary transparency allows a strong justification for unconventional alpha levels.

### Identifying Undisclosed Multiple Testing

Preregistration allows the identification of otherwise undisclosed multiple testing (e.g., Forstmeier et al., 2017). However, undisclosed multiple testing is only problematic if a researcher makes a claim based on multiple tests of the same joint hypothesis. For example, a researcher might claim that "eating jelly beans of any color is linked to acne," and then conduct multiple tests of this hypothesis by investigating whether acne is associated with eating green jelly beans, red jelly beans, yellow jelly beans, and so on, for 20 different colors of jelly bean (Munroe, 2011). In this case, if a significant association is only found for green jelly beans (e.g., $p = .023$), then it would be inappropriate for the researcher to conceal their other 19 tests and conclude that "eating jelly beans of any color is linked to acne," because the familywise error rate (the combined Type I error rate for all 20 tests of the joint hypothesis) is relevant, and the alpha level for each of the 20 tests should have been adjusted downwards (e.g., from .050 to .050/20 or .0025).

Preregistration's historical transparency allows the identification of otherwise concealed multiple testing by revealing researchers' planned multiple tests (e.g., the tests of the 20 different colors of jelly bean). However, contemporary transparency provides the same information by making research data and materials publicly available (Rubin, 2017b, pp. 272-273). For example, in the jelly beans study, a quick check of the research data and materials would show that 20 different colors of jelly beans were tested. Consequently, readers would be warranted in questioning why only green jelly beans were used to test the hypothesis that "eating jelly beans *of any color* is linked to acne."

Of course, researchers could delete any mention of non-green jelly beans from their research data and materials. However, this action would represent fraud, and fraudulent researchers may also corrupt the preregistration process. For example, they may collect data, conduct analyses, and then submit a preregistration plan and hypotheses that aligns with their results and pretend that they collected their data after registering their plan. Hence, the possibility of fraud does not help to distinguish between preregistration and non-preregistration.

It is also important to distinguish between multiple testing and multiple cases of individual testing. Multiple testing occurs when several tests are used to make a single claim about a joint hypothesis. In contrast, individual testing occurs when a single test is used to make a single claim about an individual hypothesis (Rubin, 2017b, 2019a; Tukey, 1953, pp. 82-83). Familywise error rates and alpha adjustments are only required in the case of multiple testing. They are not required in the case of individual testing, even if multiple cases of individual testing occur within the same study. This point is important because it implies that undisclosed individual testing does not inflate the alpha level for reported individual tests. For example, imagine that a researcher tests whether acne is associated with eating 20 different colors of jelly bean. If the researcher only makes claims about specific colors of jelly bean (e.g., "eating green jelly beans is linked to acne, $p = .023$"), then they are not testing a joint hypothesis, and an unadjusted conventional alpha level of .050 is appropriate in each case (Lew, 2019, pp. 21-22). Furthermore, failing to disclose some of these individual tests will not impact on the validity of claims based on other individual tests. For example, the probability statement that "eating green jelly beans is linked to acne, $p = .023$" is valid regardless of whether the tests of the other colors of jelly bean are concealed or revealed.

### Forking Paths

Forking paths refer to sample-contingent tests in which researchers make decisions about which tests to conduct based on information from their sample (Gelman & Loken, 2013, 2014; see also Rubin's, 2017a, p. 324, "result-neutral forking paths"). For example, a researcher might decide to include Variable A as a covariate in a test because it is significantly correlated with their outcome variable. In this case, a replication of their test would need to follow this same decision rule: Include Variable A as a covariate when it is correlated with the outcome variable, but exclude it as a covariate when it is not correlated with the outcome variable.

Forking paths represent a special case of the multiple testing problem, because each fork represents two potential tests of the same joint null hypothesis in a long run of hypothetical replications that follows the associated decision rule (e.g., the test that includes Covariate A and the test that excludes Covariate A). Consequently, each forking path requires an alpha adjustment (e.g., to $\alpha/2$).

Preregistration allows researchers to document forking paths ahead of time. However, forking paths do not need to be preregistered to be identified (e.g., Gelman & Loken, 2013, 2014) and accommodated (Rubin, 2017b, p. 273). For example, if a researcher explains that Variable A was included as a covariate in their test because it was significantly correlated with their outcome variable, then it is clear that they have undertaken a sample-contingent test. Consequently, it is clear that (a) future replications that find no significant correlation between Variable A and the outcome variable should not include Variable A as a covariate, and (b) the alpha level for this test should be adjusted (i.e., to $\alpha/2$) to accommodate these two potential tests.

It is also important to appreciate that the problem of forking paths may be avoided altogether if researchers and their readers condition reported $p$ values on the analytical path that is actually followed in that particular sample rather than on the two potential paths that might be followed in hypothetical replications (Cox, 1958, p. 359-361; Cox & Mayo, 2010, p. 296; Mayo, 2014, p. 232; Reid & Cox, 2015, p. 300). In other words, forking paths are not relevant in cases of *conditional inference* in which probability statements are conditioned on actual, current tests. Forking paths are only relevant in cases of *unconditional inference* in which probability statements are conditioned on both actual tests and variations of those tests that may be conducted in a long run of hypothetical replications. Hence, a researcher and their readers may make the conditional inference that a $p$ value refers to a long run of hypothetical replications of a test that *always* includes Covariate A. In this case, no alpha adjustment is required to account for the potential version of the test that does not include Covariate A.

### Identifying p-hacking

*P*-hacking refers to result-contingent tests in which researchers continue to conduct analyses until they achieve a significant result (e.g., Simmons et al., 2011). Preregistration's historical transparency may be used to identify *p*-hacking. However, historical transparency is not necessary when contemporary transparency is available. For example, researchers can confirm the absence of *p*-hacking in their research reports by (a) actively affirming the disclosure of their data collection stopping rule, data exclusions, measures, and manipulations (Simmons et al., 2012), (b) providing logical and principled justifications for nonstandard data exclusions and analytical approaches (Giner-Sorolla, 2012, p. 568), (c) providing public access to their standard data analysis procedures (e.g., Lin & Green, 2016), (d) providing public access to their research materials, data, and coding information (e.g., Aalbersberg et al., 2018), and (e) reporting the results of robustness analyses (e.g., Thabane et al., 2013).

### Identifying Optional Stopping

Preregistration's historical transparency can be used to identify otherwise undisclosed optional stopping in which researchers repeat the same test at different stages of their data collection until they obtain a significant result (e.g., Lakens, 2019). Specifically, historical transparency allows readers to check whether researchers stopped collecting data earlier, later, or at the same point at which they planned to stop collecting data. However, historical transparency is not necessary when contemporary transparency is available. In the case of contemporary transparency, researchers may actively affirm that they did not conduct any interim data analyses prior to their reported data analysis. They may also demonstrate the robustness of their results by using Bayesian hypothesis tests, because optional stopping is not problematic for Bayesian hypothesis testing (e.g., Edwards et al., 1963, p. 193; Rouder, 2014; Wagenmakers et al., 2010, p. 167).

It is also important to note that, assuming a potentially infinite number of tests, a data collection stopping rule of $p \leq \alpha$ guarantees a significant result. Hence, if researchers merely aim to decide whether or not a test result is significantly different from a point null hypothesis (in either direction), then a $p \leq \alpha$ stopping rule will result in "sampling to a forgone conclusion" (Wagenmakers, 2007, pp. 784-785). However, in many cases, researchers aim to decide whether significant results are consistent or inconsistent with directional (dividing) hypotheses (e.g., Cortina & Dunlap, 1997, p. 168; Cox, 1977, p. 51-52; Tukey, 1991, p. 100). In these cases, an optional stopping rule of "test until significant" is independent from the research conclusion because significant results may be either consistent or inconsistent with directional hypotheses. For example, a significant positive correlation is consistent with a dividing null of $r \geq 0$ but inconsistent with a dividing null of $r \leq 0$, whereas a significant negative correlation is consistent with a dividing null of $r \leq 0$ but inconsistent with a dividing null of $r \geq 0$. Consequently, a replication of the "test until significant" rule may produce a significant result that is in the opposite direction to the original result. Hence, testing until significant does not result in sampling to a forgone conclusion when the conclusion is that the significant result either supports or contradicts a directional hypothesis.

### *P Values Lose Their Meaning in Exploratory Analyses*

Some commentators have argued that preregistration is necessary because *p* values lose their meaning in exploratory (non-preregistered) analyses (e.g., Centre for Open Science, n.d.; de Groot, 2014; Nosek et al., 2018, p. 2604; Wagenmakers, 2016). According to this argument, the number of significance tests in an exploratory analysis is unplanned and sample- and result-contingent. Consequently, the number of tests in a replication of that analysis is variable and unknown. Researchers who replicate the same exploratory analysis might end up with 20 tests in some cases and 200 in others. Without knowing the precise number of tests, it is not possible to (a) calculate the familywise error rate for an exploratory analysis and (b) adjust the alpha level for each test to control this familywise error rate at its nominal level. Hence, *p* values lose their meaning in exploratory analyses because they cannot be compared to an appropriately adjusted alpha level. Preregistration solves this problem by providing a clearly defined plan of the number of tests in the data analysis.

However, the familywise error rate for the entire set of tests in an exploratory data analysis is only relevant if researchers are interested in testing a joint null hypothesis that may be rejected following at least one significant result in this analysis. In practice, researchers are unlikely to be interested in this *studywise error rate*, because the associated *studywise hypothesis* is not likely to be theoretically meaningful (Rubin, 2017a, pp. 324-325; Rubin, 2017b, 2019a).

To illustrate, consider an exploratory psychology study that measures participants' gender, age, social class, intelligence, conscientiousness, and self-esteem. In this case, a potential two-sided studywise joint null hypothesis is: "Men, older adults, and people from a lower social class background do not have either better or worse intelligence, conscientiousness, or self-esteem than women, younger adults, and people from a higher social class background, respectively." This studywise null hypothesis may be rejected if the researcher finds at least one significant gender, age, or social class difference in either intelligence, conscientiousness, or self-esteem during their analysis. Hence, each of these tests would require an alpha adjustment if the researcher wanted to maintain the studywise error rate for the studywise hypothesis at its nominal level. However, researchers are rarely interested in testing studywise hypotheses such as this, because studywise hypotheses rarely relate to meaningful theories. For example, in the current case, no extant theory assumes that gender, age, and social class differences in intelligence, conscientiousness, and self-esteem represent different instances of the same effect. Instead, each effect is likely to be driven by a different underlying process.

Hence, researchers are rarely interested in calculating studywise error rates for studywise hypotheses. Instead, as described previously, they are interested in calculating conditional familywise error rates for theoretically meaningful joint null hypotheses in which different tests are assumed to assess the same underlying effect. For example, a researcher might test the joint null hypothesis that "men do not have better self-esteem than women on either Self-Esteem Measure 1 or Self-Esteem Measure 2." In this case, the associated $p$ values do not lose their meaning in exploratory analyses, because alpha adjustments can be applied relative to the specific joint null hypothesis in question (e.g., $\alpha/2$ in the current example).

# Research Conclusions

Preregistration has been proposed as way of facilitating readers' judgments about the credibility of research conclusions. In particular, preregistration has been proposed as a method of identifying researchers' biases, selective reporting, test severity, and null results. It has also been proposed as a method of reducing publication bias and improving replication rates. I discuss each issue in turn.

## Identifying Researchers' Biases

Researchers' personal biases influence the design of preregistered tests, which influences the type of results that are possible (Donkin & Szollosi, 2020; Landy et al., 2020; Mayo, 2018, p. 10, p. 91; Silberzahn et al., 2018). Researchers' personal biases may also influence their preregistered interpretation of potential results (e.g., which type of results will be accepted as confirming or disconfirming hypotheses).

Preregistration's historical transparency allows readers to identify when researchers have changed their initial, preregistered biases to a different set of biases. However, biases do not necessarily imply errors (Kruglanski & Ajzen, 1983, pp. 19-20), and historical transparency is not necessary in order for readers to identify either biases or errors in the researchers' final conclusions. Contemporary transparency is sufficient to reveal these biases or errors.

Preregistration has also been proposed as a method of identifying when researchers fool themselves into believing their own conclusions (e.g., Field et al., 2020; Nosek et al., 2018; Wagenmakers, 2019; Wagenmakers et al., 2012). However, it is often scientifically beneficial for researchers and their readers to "fool" themselves in this way.

Confirmation bias is a tendency for researchers to interpret new evidence as confirming their hypotheses. This bias is stronger when researchers view hypotheses as being more plausible (e.g., Butzer, 2019; Hergovich et al., 2010). Hence, confirmation bias *facilitates* scientific progress by preventing well-established, highly plausible theories from being disconfirmed too easily. For example, many scientists would not regard it as "foolish" to be biased towards confirming the theory of evolution relative to creationist accounts.

Hindsight bias is a tendency for researchers to fool themselves into believing that they predicted a result after they become aware of that result. Hence, researchers and readers may fool themselves into believing that they have not changed their beliefs about a hypothesis because they "knew it all along." Importantly, however, this hindsight bias does not prevent researchers and readers from undergoing attitude change about the hypothesis in question. It only obscures this attitude change from conscious awareness (e.g., Kane et al., 2010). Hence, researchers and readers may unconsciously update their belief about a hypothesis, even if they fool themselves into believing that they have not. Again, the hindsight bias is functional because it prevents peoples' need for self-consistency from obstructing scientific progress.

## Identifying Selective Reporting

Preregistration has been proposed as a means of identifying the selective reporting of results to support certain preferred conclusions ("cherry-picking"; e.g., Nosek et al., 2018). However, selective reporting may be biased or unbiased, and only biased selective reporting is problematic. In the case of unbiased selective reporting, unreported results do not bias the final substantive research conclusion because they are unrelated to that conclusion. By analogy, imagine that a woman uses a single urine sample to conduct (a) a pregnancy test and (b) a drug test. She later reports the positive result for her pregnancy test but hides the positive result for her drug test. This selective disclosure does not affect the validity of the woman's claim that she is pregnant. In contrast, biased selective reporting would entail the woman taking 100 pregnancy tests and then only reporting a single positive result while concealing the other 99 negative results. To illustrate further, unbiased selective reporting would entail reporting all tests of gender on self-esteem and concealing all tests of age on prejudice in a research report that only makes claims about gender differences in self-esteem (see also Rubin, 2017c, p. 316).

Biased selective reporting is problematic. However, contemporary transparency is sufficient for its identification in non-preregistered research. In particular, researchers may (a) actively affirm their disclosure of all relevant variables in their research (Simmons et al., 2012), (b) provide public access to their research materials, data, and coding information, and (c) report the results of robustness analyses.

## Identifying the Severity of Tests

A test is severe if it has a low probability of confirming a hypothesis when that hypothesis is false (Mayo, 1996, 2018). For example, a severe pregnancy test would be one in which a pregnancy is only confirmed if five different pregnancy tests all yield positive results. Test severity is important because severe tests provide more credible results.

Preregistration may be used to identify the (subjectively) a priori and post hoc status of hypotheses. However, knowing that a hypothesis is a priori or post hoc does not provide any direct information about the severity with which that hypothesis has been tested. Hence, as Lakens (2019) explained, "the severity with which a claim is tested is not necessarily impacted by preregistration....[and], preregistration does not automatically increase the severity of a test" (p. 227). Indeed, preregistered a priori hypotheses can undergo *less* severe tests than non-preregistered post hoc hypotheses.

Non-preregistered statistical analyses can be severe. As Mayo (1996) explained, "violating predesignation does not necessarily conflict with NP [Neyman-Pearson] principles" (p. 295). She argued that "creative postdesignation" (i.e., "exploratory" analyses) can also be used to provide severe tests and concluded that "considerations of the creative postdesignationist provide the groundwork for justifying a break with overly narrow construals of NP methodology" (p. 318).

Non-preregistered substantive inferences can also be severe in the context of HARKing. As Mayo (2015, p. 61) explained, "what matters is not whether [hypothesis] *H* was deliberately constructed to accommodate data *x*, but how well the data, together with background information, rules out ways in which an inference to *H* can be in error" (see also Rubin, 2017c, p. 313). For example, in non-preregistered research, (a) post hoc model assumption checks can be used to select the most appropriate statistical null model for testing (Devezer et al., 2020; Spanos, 2010, p. 216), and (b) post hoc robustness analyses can be used to rule out ways in which tests of the substantive hypothesis are in error (Mayo, 2018, p. 281). Both of these post hoc approaches increase the severity of the test.

In summary, non-preregistered research can involve severe tests, and preregistration's historical transparency is not necessary to evaluate test severity. Contemporary transparency is sufficient to evaluate test severity.

### Increasing the Reporting of Null Results

Preregistration, and especially registered reports, helps to increase the reporting of null findings (e.g., Allen & Mehler, 2019; Scheel et al., 2020). However, null findings represent the absence of evidence rather than evidence of an absence (i.e., evidence of no effect; Altman & Bland, 1995). Hence, preregistration's revelation of null findings does not serve to balance positive evidence with negative evidence. Instead, it merely increases the reporting of nonevidential results.

Certainly, the reporting of null results may (a) increase the precision of estimates of effect sizes in meta-analyses, (b) prevent future researchers from conducting similarly nondiagnostic tests, and (c) hint at the presence of potential moderator variables when compared with significant results. However, none of these points mitigate the point that null results represent the absence of evidence. Consequently, expert statisticians have advised that researchers may safely "ignore" null results (Fisher, 1935, p. 16). From the perspective of hypothesis testing in basic research, it is more important to publish significant disconfirming results than to publish nonsignificant results.

### Reducing Publication Bias

Publication bias occurs when there is a preference to publish significant and confirming results over nonsignificant or disconfirming results. Preregistration's historical transparency addresses publication bias by revealing planned tests whose results are unpublished. However, contemporary transparency (i.e., publicly available data and materials) is sufficient to allow the identification and testing of any relevant but unpublished hypotheses.

Registered reports help to reduce publication bias by ensuring that editors and peer reviewers provide in principle acceptance of a research report for the publication solely on the basis of the information in its Introduction, Method, and quality check sections and not on the basis of its substantive research results (Chambers et al., 2019; Nosek & Lindsay, 2018). In this case, editors and reviewers are not influenced by whether the substantive research results are confirmatory or disconfirmatory, significant or nonsignificant.

However, preregistration is not necessary to undertake results-blind peer review (Locascio, 2017; Walster & Cleary, 1970). Editors and peer reviewers can review the Introduction, Method, and quality check sections of a non-preregistered study without being provided with access to the study's substantive results. A publication decision can then be made in the absence of knowledge about the study's results.[2] Hence, results-blind peer review is not an essential feature of preregistration, and so preregistration does not necessarily confer a credibility advantage in this respect.

**Improving Replication Rates**

Many of the issues discussed above (e.g., HARKing, multiple testing, optional stopping, *p*-hacking, selective reporting, etc.) are thought to result in artificially inflated or false positive effects that contribute to relatively low replication rates (e.g., Simmons et al., 2011). Preregistration's historical transparency makes these questionable research practices visible to readers, and so it may (a) deter researchers from engaging in some of these practices and (b) reduce the potential for the resulting research findings to be published in peer reviewed outlets. However, contemporary transparency may also achieve these outcomes.

It is also important to recognize that low replication rates have been attributed to more than just questionable research practices. For example, low replication rates have been attributed to (a) insufficiently stringent evidence thresholds (Benjamin et al., 2018), (b) insufficiently lenient evidence thresholds (Devezer et al., 2020), (c) poor measurement (e.g., Loken & Gelman, 2017), (d) model misspecification (Devezer et al., 2020), (e) low power (e.g., Rossi, 1990), (f) poor theory (e.g., Oberauer & Lewandowsky, 2019; Szollosi & Donkin, 2019), (g) an underappreciation of the influence of hidden moderators (Rubin, 2019b), and (h) errors in substantive inference (Jussim et al., 2016; Rubin, 2017b, p. 274). Hence, it is unclear whether preregistration is targeting the right set of issues to increase replication rates.

# Conclusion

Nosek et al. (2018) concluded that "preregistration improves the interpretability and credibility of research findings" (p. 2605). The present article provides a critical analysis of some of the reasons behind this conclusion.

Some of preregistration's benefits are not essential features of preregistration. In particular, planning and results-blind peer review can be achieved in non-preregistered research. Hence, preregistration is not necessary to achieve the related benefits of improved research quality and reduced publication bias.

In some cases, preregistration reveals information that does not affect the credibility of the research results. In particular, preregistration's historical transparency may reveal information about (a) HARKing, (b) irrelevant a priori hypotheses, (c) whether analyses are confirmatory or exploratory, (d) conventional alpha levels, (e) undisclosed individual testing, (f) theoretically meaningless studywise hypotheses, (g) forking paths in conditional inference, (h) optional stopping in Bayesian hypothesis tests, (i) testing until significant with directional hypotheses, (j) deviations from planned analyses, (k) confirmation and hindsight biases, (l) unbiased selective reporting, and (m) null results.

In other cases, contemporary transparency is sufficient to reveal information that does affect the credibility of the research findings, including (a) the results of relevant but unreported hypotheses, (b) unconventional alpha levels, (c) otherwise concealed multiple testing, (d) forking paths in unconditional inference, (e) *p*-hacking, (f) the absence of interim analyses in significance testing, (g) researchers' biases and errors, (h) biased selective reporting, and (i) test severity.

Finally, it is unclear whether preregistration improves replication rates. Table 1 provides a summary of the key points raised in this article.

Based on the current review, it is concluded that preregistration's historical transparency does not facilitate judgments about the credibility of research findings when contemporary transparency is available. In particular, preregistration does not facilitate judgments of credibility when researchers provide (a) clear rationales for their current hypotheses and analytical approaches, (b) public access to their research data, materials, and code, and (c) demonstrations of the robustness of their research conclusions to alternative interpretations and analytical approaches.

Of course, in many cases, researchers do not provide a sufficient degree of contemporary transparency, and in these cases preregistration's historical transparency may provide some useful information. However, historical transparency is a relatively narrow, researcher-centric form of transparency because it focuses attention on the predictions made by specific researchers at a specific time. In contrast, contemporary transparency allows research data to be interpreted and interrogated from multiple, unplanned, theoretical and analytical perspectives while maintaining a high degree of credibility. Hence, in my view, the open science movement should push more towards contemporary transparency and less towards historical transparency.

Table 1
*Summary of Key Points*

| Issue | Key Points |
|---|---|
| **Generating Hypotheses** | |
| Facilitating the development of hypotheses and tests | • Although planning and feedback may improve the quality of research, they are not essential to preregistration, and so preregistration does not necessarily confer an advantage in this respect. |
| HARKing | • Researchers can engage in transparent HARKing.<br>• Secretive HARKing does not reduce the credibility of research findings. |
| A priori hypotheses that are suppressed | • The suppression of irrelevant a priori hypotheses does not affect the credibility of research conclusions.<br>• Publicly available research data and materials allow the testing of relevant and irrelevant hypotheses. |
| **Data Analyses** | |
| Deviations from planned analyses | • We should be more interested in the rationale for the current method and analyses than in the rationale for historical changes that have led up to the current method and analyses. |
| Distinguishing confirmatory and exploratory analyses | • The distinction between confirmatory and exploratory analyses is irrelevant to the credibility of research results.<br>• Preregistration does not clarify this distinction. |
| Pre-specifying alpha levels | • Conventional alpha levels are common and do not require a justification. |
| | • Unconventional alpha levels can be justified in non-preregistered research reports. |
| Undisclosed multiple testing | • Undisclosed multiple testing can be identified by checking publicly available research data and materials. |
| | • Undisclosed individual testing does not inflate the alpha level for reported individual tests. |
| Forking paths | • Forking paths can be identified and accommodated in cases of unconditional inference in non-preregistered research. |
| | • Forking paths are not relevant in cases of conditional inference. |
| *p*-hacking | • *P*-hacking can be identified in non-preregistered research via contemporary transparency (i.e., active disclosures and justifications, publicly accessible research data and materials, robustness analyses). |
| Optional stopping | • Researchers can actively affirm that they did not conduct any interim analyses. |
| | • Robustness analyses can confirm research findings using Bayesian hypothesis tests. |
| | • Testing until significant does not result in a forgone conclusion in the case of nondirectional hypotheses. |
| *P* values in exploratory analyses | • *P* values only lose their meaning in exploratory analyses if researchers test studywise hypotheses. |
| | • Researchers do not usually test studywise hypotheses. |

### Research Conclusions

| | |
|---|---|
| Researchers' biases | • Biases do not necessarily imply errors, and both biases and errors can be identified in non-preregistered research. |
| | • Confirmation and hindsight biases can facilitate scientific progress. |
| Selective reporting | • Unbiased selective reporting is not problematic. |
| | • Biased selective reporting can be identified via contemporary transparency. |
| Test severity | • Non-preregistered tests can be severe. |
| | • Contemporary transparency is sufficient to evaluate test severity. |
| Reporting null results | • Null findings represent the absence of evidence. |
| | • It is more important to publish significant disconfirming results than to publish nonsignificant results. |
| Publication bias | • Contemporary transparency is sufficient to identify and test any relevant unpublished hypotheses. |
| | • Results-blind peer review is not an essential feature of preregistration. |
| Replication rates | • Contemporary transparency may deter questionable research practices. |
| | • Low replication rates may be caused by issues that are unrelated to questionable research practices. |

*Note.* Contemporary transparency refers to (a) clear rationales for current hypotheses and analytical approaches, (b) public access to research data, materials, and code, and (c) demonstrations of the robustness of research conclusions to alternative interpretations and analytical approaches.

# References

Aalbersberg, I. J., Appleyard, T., Brookhart, S., Carpenter, T., Clarke, M., Curry, S., … Vazire, S. (2018, February 15). *Making science transparent by default: Introducing the TOP statement.* https://doi.org/10.31219/osf.io/sm78t

Allen, C., & Mehler, D. M. (2019). Open science challenges, benefits and tips in early career and beyond. *PLoS Biology, 17*(5), e3000246. https://doi.org/10.1371/journal.pbio.3000246

Altman, D. G., & Bland, J. M. (1995). Absence of evidence is not evidence of absence. *BMJ, 311*(7003), 485. https://doi.org/10.1136/bmj.311.7003.485

Baron, J. (2018, May 10). *Prediction, accommodation and pre-registration.* http://judgmentmisguided.blogspot.com/2018/05/prediction-accommodation-and-pre.html?m=1

Benjamin, D. J., Berger, J. O., Johannesson, M., Nosek, B. A., Wagenmakers, E. J., Berk, R.,...& Cesarini, D. (2018). Redefine statistical significance. *Nature Human Behaviour, 2*(1), 6-10. https://doi.org/10.1038/s41562-017-0189-z

Bleidorn, W., Arslan, R. C., Denissen, J. J., Rentfrow, P. J., Gebauer, J. E., Potter, J., & Gosling, S. D. (2016). Age and gender differences in self-esteem: A cross-cultural window. *Journal of Personality and Social Psychology, 111*(3), 396-410. http://dx.doi.org/10.1037/pspp0000078

Butzer, B. (2019). Bias in the evaluation of psychology studies: A comparison of parapsychology versus neuroscience. *EXPLORE: The Journal of Science & Healing.* https://doi.org/10.1016/j.explore.2019.12.010

Centre for Open Science. (n.d.). *What is preregistration?* https://cos.io/prereg/

Chambers, C. D., Forstmann, B., & Pruszynski, J. A. (2019). Science in flux: Registered Reports and beyond at the European Journal of Neuroscience. *European Journal of Neuroscience, 49*(1), 4-5. http://dx.doi.org/10.1111/ejn.14319

Coffman, L. C., & Niederle, M. (2015). Pre-analysis plans have limited upside, especially where replications are feasible. *Journal of Economic Perspectives, 29,* 81–98. http://dx.doi.org/10.1257/jep.29.3.81

Cortina, J. M., & Dunlap, W. P. (1997). On the logic and purpose of significance testing. *Psychological Methods, 2,* 61-172. http://doi.org/10.1037/1082-989X.2.2.161

Cox, D. R. (1958). Some problems connected with statistical inference. *Annals of Mathematical Statistics, 29*(2), 357-372. http://dx.doi.org/10.1214/aoms/1177706618

Cox, D. R. (1977). The role of significance tests. *Scandinavian Journal of Statistics, 4,* 49-70. https://www.jstor.org/stable/4615652

Cox, D. R., & Mayo, D. G. (2010). Objectivity and conditionality in frequentist inference. In D. G. Mayo & A. Spanos (Eds.), *Error and inference: Recent exchanges on experimental reasoning, reliability, and the objectivity and rationality of science* (pp. 276-304). Cambridge University Press.

de Groot, A. D. (2014). The meaning of "significance" for different types of research (E. J. Wagenmakers, D. Borsboom, J. Verhagen, R. Kievit, M. Bakker, A. Cramer, . . . H. L. J. van der Maas). *Acta Psychologica, 148,* 188–194. http://dx.doi.org/10.1016/j.actpsy.2014.02.001

Devezer, B., Navarro, D. J., Vandekerckhove, J., & Buzbas, E. O. (2020, April 28). *The case for formal methodology in scientific reform.* https://doi.org/10.1101/2020.04.26.048306

Donkin, C., & Szollosi, A. (2020, January 2). Unpacking the disagreement: Guest post by Donkin and Szollosi. *Bayesian Spectacles.* http://www.bayesianspectacles.org/unpacking-the-disagreement-guest-post-by-donkin-and-szollosi/

Edwards, W., Lindman, H., & Savage, L. J. (1963). Bayesian statistical inference for psychological research. *Psychological Review, 70*(3), 193-242. https://doi.org/10.1037/h0044139

Field, S. M., Wagenmakers, E. J., Kiers, H. A., Hoekstra, R., Ernst, A. F., & van Ravenzwaaij, D. (2020). The effect of preregistration on trust in empirical research findings: Results of a registered report. *Royal Society Open Science, 7*(4), 181351. https://doi.org/10.1098/rsos.181351

Feynman, R. (1974). *Cargo cult science.* Caltech's 1974 commencement address. http://calteches.library.caltech.edu/51/2/CargoCult.htm

Fisher, R. A. (1935). *The design of experiments*. Oliver & Boyd.

Forstmeier, W., Wagenmakers, E. J., & Parker, T. H. (2017). Detecting and avoiding likely false-positive findings–a practical guide. *Biological Reviews, 92,* 1941-1968. http://dx.doi.org/10.1111/brv.12315

Gelman, A., & Loken, E. (2013). *The garden of forking paths: Why multiple comparisons can be a problem, even when there is no "fishing expedition" or "p-hacking" and the research hypothesis was posited ahead of time.* Department of Statistics, Columbia University. http://www.stat.columbia.edu/~gelman/research/unpublished/p_hacking.pdf

Gelman, A., & Loken, E. (2014). The statistical crisis in science. *American Scientist, 102,* 460. http://dx.doi.org/10.1511/2014.111.460

Giner-Sorolla, R. (2012). Science or art? How aesthetic standards grease the way through the publication bottleneck but undermine science. *Perspectives on Psychological Science, 7,* 562-571. http://dx.doi.org/10.1177/1745691612457576

Hergovich, A., Schott, R., & Burger, C. (2010). Biased evaluation of abstracts depending on topic and conclusion: Further evidence of a confirmation bias within scientific psychology. *Current Psychology, 29*(3), 188-209. https://doi.org/10.1007/s12144-010-9087-5

Hollenbeck, J. R., & Wright, P. M. (2017). Harking, sharking, and tharking: Making the case for post hoc analysis of scientific data. *Journal of Management, 43*(1), 5-8. https://doi.org/10.1177/0149206316679487

Jussim, L., Crawford, J. T., Anglin, S. M., Stevens, S. T., & Duarte, J. L. (2016). Interpretations and methods: Towards a more effectively self-correcting social psychology. *Journal of Experimental Social Psychology, 66,* 116-133. https://doi.org/10.1016/j.jesp.2015.10.003

Kane, M. J., Core, T. J., & Hunt, R. R. (2010). Bias versus bias: Harnessing hindsight to reveal paranormal belief change beyond demand characteristics. *Psychonomic Bulletin & Review, 17*(2), 206-212. https://doi.org/10.3758/PBR.17.2.206

Kerr, N. L. (1998). HARKing: Hypothesizing after the results are known. *Personality and Social Psychology Review, 2,* 196-217. http://dx.doi.org/10.1207/s15327957pspr0203_4

Kruglanski, A. W., & Ajzen, I. (1983). Bias and error in human judgment. *European Journal of Social Psychology, 13*(1), 1-44. https://doi.org/10.1002/ejsp.2420130102

Lakens, D. (2019). *The value of preregistration for psychological science: A conceptual analysis*. *Japanese Psychological Review, 62*(3), 221-230.

Landy, J. F., Jia, M. (L.), Ding, I. L., Viganola, D., Tierney, W., Dreber, A., Johannesson, M., Pfeiffer, T., Ebersole, C. R., Gronau, Q. F., Ly, A., van den Bergh, D., Marsman, M., Derks, K., Wagenmakers, E.-J., Proctor, A., Bartels, D. M., Bauman, C. W., Brady, W. J.,…The Crowdsourcing Hypothesis Tests Collaboration. (2020). Crowdsourcing hypothesis tests:

Making transparent how design choices shape research results. *Psychological Bulletin.* https://doi.org/10.1037/bul0000220

Ledgerwood, A. (2018). The preregistration revolution needs to distinguish between predictions and analyses. *Proceedings of the National Academy of Sciences, 115,* E10516-E10517. http://dx.doi.org/10.1073/pnas.1812592115

Leung, K. (2011). Presenting post hoc hypotheses as a priori: Ethical and theoretical issues. *Management and Organization Review, 7,* 471–479. http://dx.doi.org/10.1017/CBO9781139171434.009

Lew, M. J. (2019). A reckless guide to *p*-values: Local evidence, global errors. In A. Bespalov, M. C. Michel, & T. Steckler (Eds.), *Good research practice in experimental pharmacology*. Springer. https://arxiv.org/abs/1910.02042

Lewandowsky, S. (2019, January 22). Avoiding Nimitz Hill with more than a Little Red Book: Summing up #PSpmereg. *Psychonomic Society*. https://featuredcontent.psychonomic.org/avoiding-nimitz-hill-with-more-than-a-little-red-book-summing-up-pspmereg/

Lin, W., & Green, D. P. (2016). Standard operating procedures: A safety net for pre-analysis plans. *Political Science & Politics, 49,* 495-500. http://dx.doi.org/10.1017/S1049096516000810

Locascio, J. J. (2017). Results blind science publishing. *Basic and Applied Social Psychology, 39,* 239-246. http://dx.doi.org/10.1080/01973533.2017.1336093

Loken, E., & Gelman, A. (2017). Measurement error and the replication crisis. *Science 355,* 6325, 584-585. http://dx.doi.org/10.1126/science.aal3618

Mayo, D. G. (1996). *Error and the growth of experimental knowledge.* Chicago University Press.

Mayo, D. G. (2014). On the Birnbaum argument for the strong likelihood principle. *Statistical Science, 29,* 227-239. http://dx.doi.org/10.1214/1 3-STS457

Mayo, D. G. (2015). Learning from error: How experiment gets a life (of its own). In M. Boumans, G. Hon, & A. C. Petersen (Eds.), *Error and uncertainty in scientific practice* (pp. 57-78). Routledge.

Mayo, D. G. (2018). *Statistical inference as severe testing.* Cambridge University Press.

Munroe, R. (06/04/2011). *Significant*. https://xkcd.com/882/

Navarro, D. (2019, November 9). *Paths in strange places, part I.* https://djnavarro.net/post/paths-in-strange-spaces/

Nosek, B. A., Beck, E. D., Campbell, L., Flake, J. K., Hardwicke, T. E., Mellor, D. T., ... & Vazire, S. (2019). Preregistration is hard, and worthwhile. *Trends in Cognitive Sciences, 23*(10), 815-818. http://dx.doi.org/10.1016/j.tics.2019.07.009

Nosek, B. A., Ebersole, C. R., DeHaven, A. C., & Mellor, D. T. (2018). The preregistration revolution. *Proceedings of the National Academy of Sciences, 115,* 2600-2606. http://dx.doi.org/10.1073/pnas.1708274114

Nosek, B. A., & Lindsay, D. S. (2018, March). Preregistration becoming the norm in psychological science. *APS Observer.* https://www.psychologicalscience.org/observer/preregistration-becoming-the-norm-in-psychological-science/comment-page-1

Oberauer, K., & Lewandowsky, S. (2019). Addressing the theory crisis in psychology. *Psychonomic Bulletin & Review, 26,* 1596–1618. https://doi.org/10.3758/s13423-019-01645-2

Reid, N., & Cox, D. R. (2015). On some principles of statistical inference. *International Statistical Review, 83,* 293-308. http://dx.doi.org/10.1111/insr.12067

Rossi, J. S. (1990). Statistical power of psychological research: What have we gained in 20 years? *Journal of Consulting and Clinical Psychology*, *58*(5), 646-656. http://dx.doi.org/10.1037//0022-006x.58.5.646

Rouder, J. N. (2014). Optional stopping: No problem for Bayesians. *Psychonomic Bulletin & Review, 21*(2), 301-308. https://doi.org/10.3758/s13423-014-0595-4

Rubin, M. (2017a). An evaluation of four solutions to the forking paths problem: Adjusted alpha, preregistration, sensitivity analyses, and abandoning the Neyman-Pearson approach. *Review of General Psychology, 21,* 321-329. http://dx.doi.org/10.1037/gpr0000135

Rubin, M. (2017b). Do *p* values lose their meaning in exploratory analyses? It depends how you define the familywise error rate. *Review of General Psychology, 21,* 269-275. http://dx.doi.org/10.1037/gpr0000123

Rubin, M. (2017c). When does HARKing hurt? Identifying when different types of undisclosed post hoc hypothesizing harm scientific progress. *Review of General Psychology, 21,* 308-320. http://dx.doi.org/10.1037/gpr0000128

Rubin, M. (2019a). The costs of HARKing. *The British Journal for the Philosophy of Science.* https://doi.org/10.1093/bjps/axz050

Rubin, M. (2019b). What type of Type I error? Contrasting the Neyman-Pearson and Fisherian approaches in the context of exact and direct replications. *Synthese.* https://doi.org/10.1007/s11229-019-02433-0

Scheel, A. M., Schijen, M., & Lakens, D. (2020, February 5). *An excess of positive results: Comparing the standard psychology literature with registered reports.* https://doi.org/10.31234/osf.io/p6e9c

Silberzahn, R., Uhlmann, E. L., Martin, D. P., Anselmi, P., Aust, F., Awtrey, E., ... & Carlsson, R. (2018). Many analysts, one data set: Making transparent how variations in analytic choices affect results. *Advances in Methods and Practices in Psychological Science*, *1*, 337-356. http://dx.doi.org/10.1177/2515245917747646

Simmons, J. P., Nelson, L. D., & Simonsohn, U. (2011). False-positive psychology: Undisclosed flexibility in data collection and analysis allows presenting anything as significant. *Psychological Science, 22,* 1359–1366. http://dx.doi.org/10.1177/0956797611417632

Simmons, J. P., Nelson, L. D., & Simonsohn, U. (2012). A 21 word solution. *Dialogue, 21,* 4-7. http://dx.doi.org/10.2139/ssrn.2160588

Spanos, A. (2010). Akaike-type criteria and the reliability of inference: Model selection versus statistical model specification. *Journal of Econometrics, 158*(2), 204-220. https://doi.org/10.1016/j.jeconom.2010.01.011

Szollosi, A., & Donkin, C. (2019). Arrested theory development: The misguided distinction between exploratory and confirmatory research. *PsyArxiv.* https://psyarxiv.com/suzej/

Szollosi, A., Kellen, D., Navarro, D. J., Shiffrin, R., van Rooij, I., Van Zandt, T., & Donkin, C. (2020). Is preregistration worthwhile? *Trends in Cognitive Science, 24*(2), 94-95. https://doi.org/10.1016/j.tics.2019.11.009

Thabane, L., Mbuagbaw, L., Zhang, S., Samaan, Z., Marcucci, M., Ye, C.,…Goldsmith, C. H. (2013). A tutorial on sensitivity analyses in clinical trials: The what, why, when and how. *BMC Medical Research Methodology,13,* 92. http://dx.doi.org/10.1186/1471-2288-13-92

Tukey, J. W. (1953). *The problem of multiple comparisons.* Princeton University.

Tukey, J. W. (1991). The philosophy of multiple comparisons. *Statistical Science, 6,* 100-116. http://doi.org/10.1214/ss/1177011945

Vancouver, J. B. (2018). In defense of HARKing. *Industrial and Organizational Psychology, 11,* 73-80. http://dx.doi.org/10.1017/iop.2017.89

Wagenmakers, E. J. (2007). A practical solution to the pervasive problems of *p* values. *Psychonomic Bulletin & Review, 14*(5), 779-804. https://doi.org/10.3758/BF03194105

Wagenmakers, E. J. (2016). [Comment]. https://www.psychologicalscience.org/observer/why-preregistration-makes-me-nervous

Wagenmakers, E. J. (2019, November 5). *A breakdown of "preregistration is redundant, at best.* Bayesian Spectacles. https://www.bayesianspectacles.org/a-breakdown-of-preregistration-is-redundant-at-best/

Wagenmakers, E. J., Lodewyckx, T., Kuriyal, H., & Grasman, R. (2010). Bayesian hypothesis testing for psychologists: A tutorial on the Savage–Dickey method. *Cognitive Psychology*, *60*(3), 158-189. https://doi.org/10.1016/j.cogpsych.2009.12.001

Wagenmakers, E. J., Wetzels, R., Borsboom, D., van der Maas, H. L., & Kievit, R. A. (2012). An agenda for purely confirmatory research. *Perspectives on Psychological Science, 7*(6), 632-638. https://doi.org/10.1177/1745691612463078

Walster, G. W., & Cleary, T. A. (1970). A proposal for a new editorial policy in the social sciences. *The American Statistician, 24*(2), 16-19. https://doi.org/10.1080/00031305.1970.10478884

Worrall, J. (2014). Prediction and accommodation revisited. *Studies in History and Philosophy of Science, 45,* 54–61. http://dx.doi.org/10.1016/j.shpsa.2013.10.001

## Endnotes

1. A distinction can be drawn between the *perceived* credibility of a preregistered study based on heuristics and social norms, and the *actual* credibility of a preregistered study, based on the actual impact of questionable research practices in that study. Following Field et al. (2020), I assume that "the actual credibility of a study is ultimately more important than how it is perceived" (p. 3). However, I would note that there is currently no clear evidence that preregistration increases the perceived credibility of research (Field et al., 2020).

2. For a list of journals that are currently using a results-blind peer reviewing approach, please see https://jbp.uncc.edu/other-journals-involved-in-this-joint-initiative/

## Funding

The author declares no funding sources.

## Conflict of Interest

The author declares no conflict of interest.